\documentclass[preprint,12pt]{elsarticle}
\RequirePackage[colorlinks,citecolor=blue,urlcolor=blue,linkcolor=blue]{hyperref}
\usepackage[colorlinks]{hyperref}
\usepackage[T1]{fontenc}
\usepackage[utf8]{inputenc}

\usepackage{amssymb}

\journal{Applied Radiation and Isotopes}

\begin{document}

\begin{frontmatter}



\title{Characterization of a kg-scale archaeological lead-based PbWO$_4$ cryogenic detector for the RES-NOVA experiment}


\author[LBL]{J.W.~Beeman}
\author[LNGS]{G.~Benato}
\author[LNGS]{C.~Bucci}
\author[MPI]{L.~Canonica}
\author[UNIMIB,INFN-MIB]{P.~Carniti}
\author[LNGS,GSSI]{E.~Celi}
\author[INFN-MIB]{M.~Clemenza}
\author[LNGS]{A.~D'Addabbo}
\author[INR]{F.A.~Danevich}
\author[Ge]{S.~Di Domizio}
\author[LNGS]{S.~Di~Lorenzo}
\author[ISMA]{O.M.~Dubovik}
\author[MPI]{N.~Ferreiro Iachellini}
\author[GSSI,Roma1]{F.~Ferroni}
\author[UNIMIB,INFN-MIB]{E.~Fiorini}
\author[LNGS]{S.~Fu}
\author[MPI]{A.~Garai}
\author[LNGS,GSSI]{S.~Ghislandi}
\author[UNIMIB,INFN-MIB]{L.~Gironi}
\author[LNGS]{P.~Gorla}
\author[UNIMIB,INFN-MIB]{C.~Gotti}
\author[LNGS]{P.V.~Guillaumon}
\author[LNGS,GSSI]{D.L.~Helis}
\author[NSC]{G.P.~Kovtun}
\author[MPI]{M.~Mancuso}
\author[LNGS,GSSI]{L.~Marini}
\author[LNGS]{M.~Olmi}
\author[LNGS,GSSI]{L.~Pagnanini}
\author[1,LNGS,TUM]{L.~Pattavina}
\author[INFN-MIB]{G.~Pessina}
\author[MPI]{F.~Petricca}
\author[LNGS]{S.~Pirro}
\author[UNIMIB,INFN-MIB]{S.~Pozzi}
\author[LNGS,GSSI]{A.~Puiu}
\author[LNGS,GSSI]{S.~Quitadamo\corref{cor1}\footnote{\texttt{simone.quitadamo@gssi.it, luca.pattavina@lngs.infn.it}}}
\author[TUM]{J.~Rothe}
\author[NSC]{A.P.~Scherban}
\author[TUM]{S.~Sch{\"o}nert}
\author[NSC]{D.A.~Solopikhin}
\author[TUM]{R.~Strauss}
\author[INFN-MIB]{E.~Tarabini}
\author[INR]{V.I.~Tretyak}
\author[ISMA]{I.A.~Tupitsyna}
\author[TUM]{V.~Wagner}

\affiliation[LBL]{organization={Lawrence Berkeley National Laboratory},
            city={Berkeley},
            postcode={94720}, 
            state={USA},
            country={California}}

\affiliation[LNGS]{organization={Laboratori Nazionali del Gran Sasso},
            addressline={Via G. Acitelli 22}, 
            city={Assergi},
            postcode={67100}, 
            state={IT},
            country={Italy}}

\affiliation[MPI]{organization={Max-Planck-Institut f{\"u}r Physik},
            addressline={F{\"o}hringer Ring 6}, 
            city={M{\"u}nchen},
            postcode={DE-80805}, 
            country={Germany}}

\affiliation[UNIMIB]{organization={Dipartimento di Fisica, Universit\`a di Milano - Bicocca},
            addressline={Piazza della Scienza 3}, 
            city={Milano},
            postcode={I-20126}, 
            state={IT},
            country={Italy}}

\affiliation[INFN-MIB]{organization={INFN Sezione di Milano - Bicocca},
            addressline={Piazza della Scienza 3}, 
            city={Milano},
            postcode={I-20126}, 
            state={IT},
            country={Italy}}

\affiliation[GSSI]{organization={Gran Sasso Science Institute},
            addressline={Viale F. Crespi 7}, 
            city={L'Aquila},
            postcode={67100}, 
            state={IT},
            country={Italy}}

\affiliation[INR]{organization={Institute for Nuclear Research of NASU},
            city={Kyiv},
            postcode={03028}, 
            country={Ukraine}}
            
\affiliation[Ge]{organization={INFN Sezione di Genova and Universit\`a di Genova},
            addressline={Via Dodecaneso 33}, 
            city={Genova},
            postcode={I-16146}, 
            state={IT},
            country={Italy}}
            
\affiliation[ISMA]{organization={Institute of Scintillation Materials of NASU},
            city={Kharkiv},
            postcode={61072}, 
            country={Ukraine}}

\affiliation[Roma1]{organization={INFN Sezione di Roma-1},
            addressline={P.le Aldo Moro 2}, 
            city={Roma},
            postcode={I-00185}, 
            state={IT},
            country={Italy}}

\affiliation[NSC]{organization={National Science Center 'Kharkiv Institute of Physics and Technology'},
            city={Kharkiv},
            postcode={61108}, 
            country={Ukraine}}
 
            
\affiliation[TUM]{organization={Technical University of Munich},
            addressline={James­Franck­Strasse 1}, 
            city={Garching},
            postcode={85748}, 
            state={DE},
            country={Germany}}


\begin{abstract}
Core-collapse Supernovae (SNe) are one of the most energetic events in the Universe, during which almost all the star's binding energy is released in the form of neutrinos. These particles are direct probes of the processes occurring in the stellar core and provide unique insights into the gravitational collapse. RES-NOVA will revolutionize how we detect neutrinos from astrophysical sources, by deploying the first ton-scale array of cryogenic detectors made from archaeological lead. Pb offers the highest neutrino interaction cross-section via coherent elastic neutrino-nucleus scattering (CE$\nu$NS). Such process will enable RES-NOVA to be equally sensitive to all neutrino flavors. For the first time, we propose the use archaeological Pb as sensitive target material in order to achieve an ultra-low background level in the region of interest (\textit{O}(1keV)). All these features make possible the deployment of the first cm-scale neutrino telescope for the investigation of astrophysical sources.
In this contribution, we will characterize the radiopurity level and the performance of a small-scale proof-of-principle detector of RES-NOVA, consisting in a PbWO$_4$ crystal made from archaeological-Pb operated as cryogenic detector. 
\end{abstract}


\end{frontmatter}


\section{Introduction}
The question of why and how massive stars explode in the bright cosmic fireworks known as supernovae (SNe) is one of the most fundamental and long-standing unsolved puzzles in stellar astrophysics. Developing a better understanding and eventually solving this conundrum is of utmost importance for a multitude of problems, such as: how do the properties of a SN depend on its progenitor star characteristics? Which stars collapse into neutron stars and which give birth to black holes? What is the role of SNe in the nucleosynthesis of heavy and light elements? Answering these questions will be intimately linked to progress in deciphering the physical mechanisms that cause the catastrophic collapse of a stellar core to become a SN breakout, which brings to an end the life of a massive star.

\begin{figure}[t]
\centering
\includegraphics[width=0.5\textwidth]{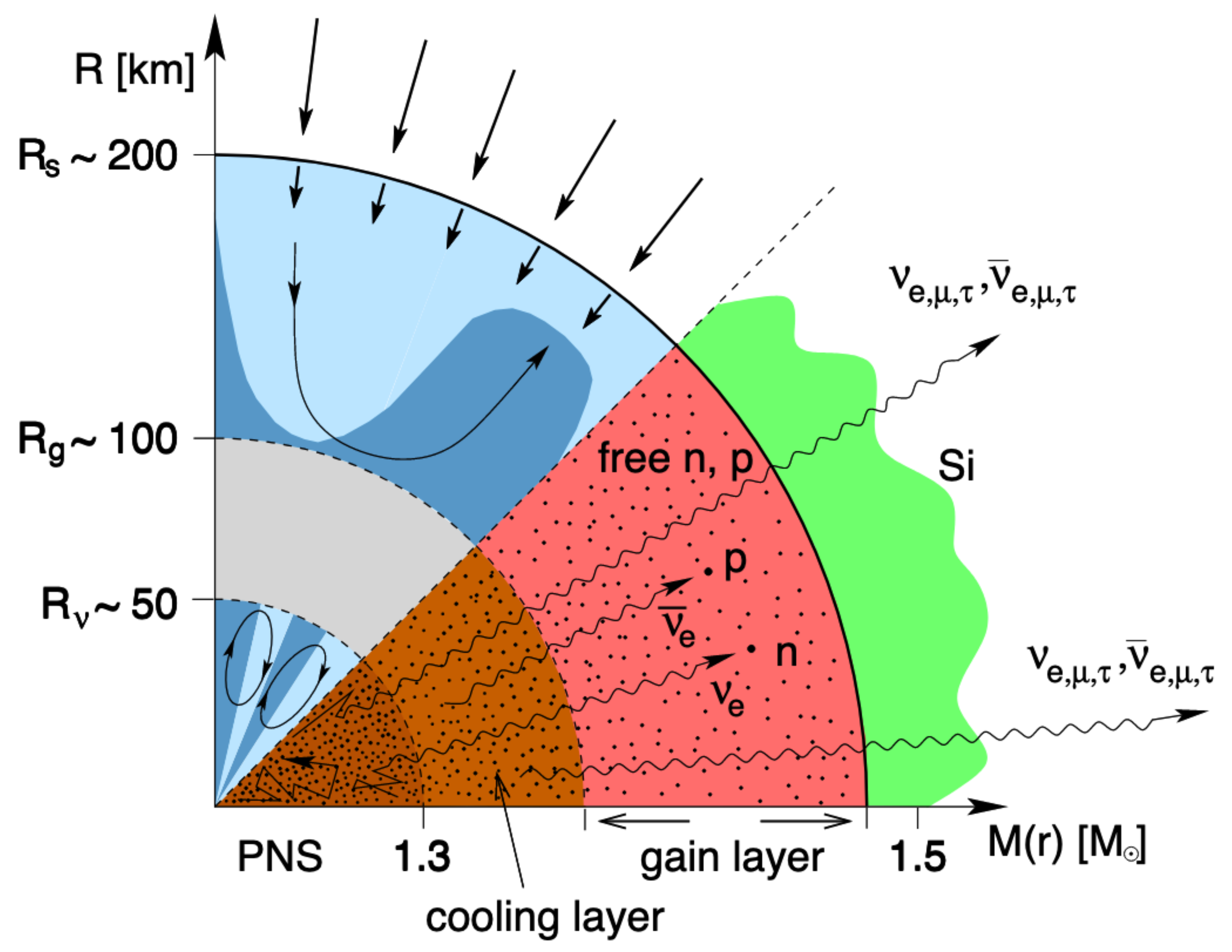}
\caption{Schematic view of a core-collapse SN during the time of its accretion when $\nu_e$/$\bar\nu_{e}$ neutrinos deposit energy in the gain region below the stalling shock wave, here at R$_s$ = 200~km. \textcolor{black}{List of labels: PNS = proto-neutron star, M(r) = mass enclosed within a radius r (in solar masses), R$_{\nu}$ = neutrinosphere radius, R$_g$ = gain-region radius, R$_s$ = shock-region radius, Si = silicon-dominated region.} Figure adapted from~\cite{Janka:2006fh}.}
\label{fig:SNcore} 
\end{figure}

Most of the energy released during the core collapse of a growing massive star is in the form of neutrinos of all flavors. Their mean free path can varies from a few meters in the core, during the formation of the protoneutron star (PNS), to free streaming in the outer stellar envelopes during the stellar outburst~\cite{Mirizzi:2015eza}. In Fig.~\ref{fig:SNcore} a simplified scheme of the SN core structure is shown, representing the time when neutrinos deposit energy in the stellar envelop, igniting the explosion. This "neutrino mechanism" is the most plausible paradigm for core-collapse SN explosions~\cite{Janka:2006fh}. The crucial role of neutrinos is explained by the fact that their mean free path is essentially comparable to the geometric size of the object~\cite{Janka:2006fh}. Photons interact about 20 orders of magnitude stronger than neutrinos, thus they have a very short mean free path, and therefore get stuck inside the stellar envelope~\cite{Mirizzi:2015eza}. On the other hand, gravitons \textcolor{black}{(the hypothetical particle that mediates the gravitational interaction)} are expected to interact so weakly that they can escape freely and drain very little energy from a SN core.

Neutrinos can be used as direct probes of the central engine of a core-collapse SN and they can also carry imprints of the astrophysical processes occurring in the explosion. Detecting these elusive particles is a game changer in understanding the physics of core-collapse and the role that neutrinos play during this event. \textcolor{black}{This was demonstrated by the Kaiomkande-II~\cite{Kamiokande1987} IMB~\cite{IMB1987} and Baksan~\cite{BAKSAN1987} experiments that in 1987 detected the first (and only) neutrinos from a SN, SN1987A. In fact, our current understanding of the explosion mechanism of SNe mostly rely on the 25 events detected by these experiments~\cite{Review1987neutrino}.}

The possibility to experimentally test and validate the models describing the processes that are steering the SN neutrino emission goes through the difficult path of detecting neutrinos of all flavors. This aspect is relevant because of uncertainties connected to neutrino flavor oscillations in the stellar medium and on Earth must be overcome. Currently running and planned neutrino telescopes are exploiting \textit{kton}-scale detectors containing either liquid scintillators, like SNO+~\cite{Rumleskie:2020iip} and the upcoming JUNO~\cite{An:2015jdp}, or water, as Super-Kamiokande~\cite{Simpson:2019xwo} or its upgrade Hyper-Kamiokande~\cite{Hyper-Kamiokande:2022smq}. These detectors are mostly sensitive to $\bar\nu_{e}$ through the inverse-beta decay (IBD) channel, which has cross-section at the level of 10$^{-41}$~cm$^2$ for 10~MeV neutrinos. A few percent of the detectable neutrino signal is attributed to $\nu_x$/$\bar\nu_{x}$ (with $x=\mu,\tau$), which can be observed via the neutrino-electron scattering neutral current processes, depending on the detector energy threshold. This has a cross-section about 2 orders of magnitude lower than IBD~\cite{Scholberg:2012id}. 

The recent discovery of the coherent elastic neutrino-nucleus scattering (CE$\nu$NS)~\cite{Akimov:2017ade} has open a wealth of opportunity in neutrino detection. This is a neutral current process, thus equally sensitive to all neutrino flavors, it has a cross-section more than a factor 10$^3$ higher than IBD~\cite{Scholberg:2012id}, and it has no kinematic threshold~\cite{Freedman:1973yd}. All these features make CE$\nu$NS an ideal channel for the detection of neutrinos with relatively small-scale detectors~\cite{Drukier:1983gj}. The RES-NOVA experiment is aiming at detecting astrophysical neutrino sources, as SNe, with a compact detector made of archaeological Pb-based cryogenic detectors, using  CE$\nu$NS as detection channel~\cite{Pattavina:2020cqc}. Pb is a unique target material, since it is the only element that simultaneously offers the highest neutrino interaction cross-section via CE$\nu$NS~\cite{Pattavina:2020cqc} and the highest nuclear stability~\cite{Beeman:2012wz}. The high neutrino interaction cross-section ($\mathcal{O}(10^{-38}~cm^2$) allows to have a high interaction rates, while the high nuclear stability helps in reaching low background levels. Then, when archaeological Pb is employed for the detector realization, even lower background level can be achieved.

\section{Detection of Supernova neutrinos via coherent elastic neutrino-nucleus scattering}
\label{sec:SN}

\begin{figure}[t]
\centering
\includegraphics[width=0.9\textwidth]{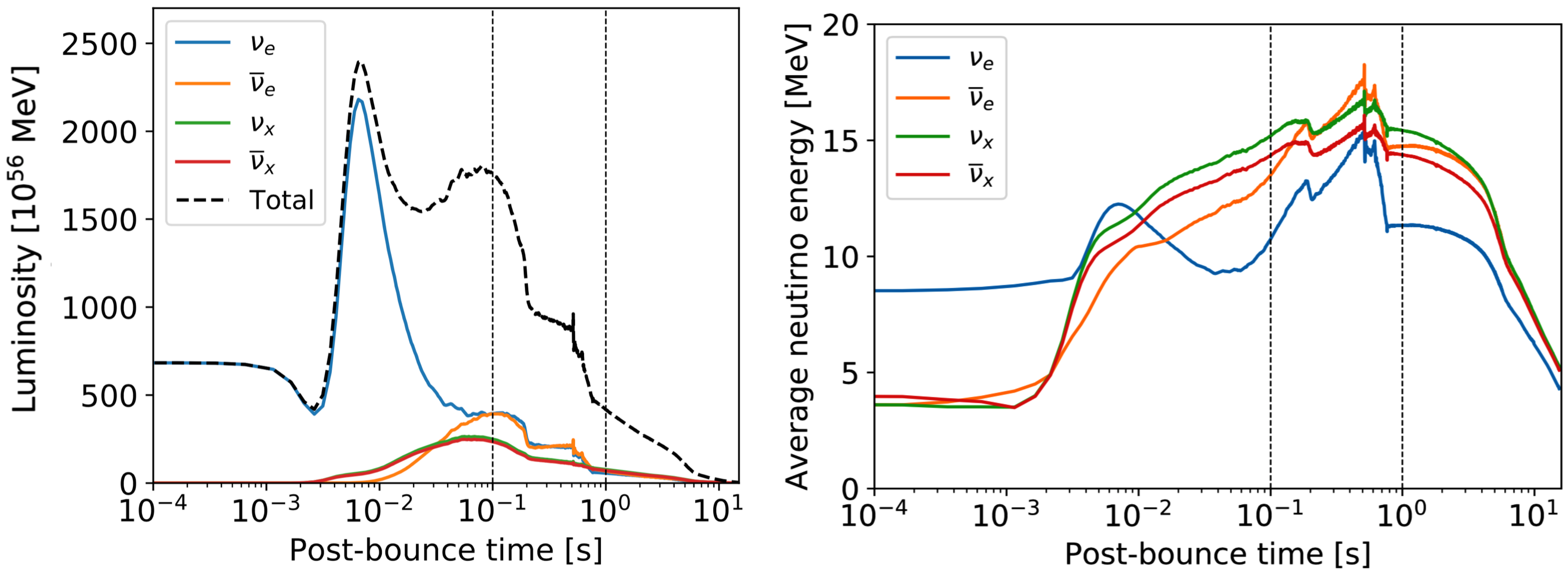}
\caption{\textcolor{black}{Luminosity and average energy of neutrinos of different flavours as a function of the post-bounce time for a 27~$M_\odot$ progenitor with equation of state (EoS) from~\cite{Lattimer:1991nc} at 10~kpc. The labels $\nu_x$,$\bar{\nu}_x$ refer to $\nu_{\mu}$,$\bar{\nu}_{\mu}$,$\nu_{\tau}$,$\bar{\nu}_{\tau}$. For details on the simulation refer to~\cite{Mirizzi:2015eza}}.}
\label{fig:SNsignal} 
\end{figure}

Neutrinos emitted from a SN are expected to have an almost thermal distribution, whose average energy increase as the time to the stellar explosion approaches. In Fig.~\ref{fig:SNsignal}, the main features of a SN neutrino emission from a 1D hydrodynamic simulation~\cite{Mirizzi:2015eza} for a successful explosion of a 27~$M_{\odot}$ progenitor star with equation of state from \cite{Lattimer:1991nc} occurring at a distance of 10~kpc is shown.

The main characteristics of a SN neutrino signal are the high signal intensity in a time scale of few seconds, an average neutrino energy of about 20~MeV and a full-flavor emission.
\textcolor{black}{The RES-NOVA experiment will detect the full-composite SN neutrino emission via the CE$\nu$NS channel. This neutral current process is highly and equally sensitive to all neutrino flavors, unlike, the currently employed IBD and  electron-scattering.}

\textcolor{black}{The differential neutrino interaction cross-section, assuming a target 0-spin nucleus and no beyond Standard Model interactions,} can be derived from basic Standard Model principles~\cite{Freedman:1973yd}:
\begin{equation}
\label{eq:xsec}
\frac{d\sigma}{d E_R} = \frac{G^2_F}{8 \pi} m_N \left[N - Z (1-4\sin^2 \theta_W)\right]^2 \left(2- \frac{E_R m_N}{E^2}\right) \cdot |F(q^2)|^2\ ,
\end{equation}
where $G_F$ is the Fermi coupling constant, $\theta_W$ the Weinberg angle, $Z$ and $N$ the atomic and neutron numbers of the target nucleus, while $m_N$ its mass, $E$ the  energy of the incoming neutrino and $E_R$ the recoil energy of the target. The last term of the equation, $F(q)$, is the elastic nuclear form factor at momentum transfer $q=\sqrt{2E_R m_N}$, and for small momentum transfers its value is close to unity. Pb is a very good target for the detection of neutrinos via CE$\nu$NS thanks to the $N^2$ dependence of the cross-section. The interaction of neutrinos with Pb can be considered as coherent for neutrino energies up to 30~MeV \cite{Drukier:1983gj}.
When SN neutrinos scatter off Pb nuclei, they induce low-energy nuclear recoils with energies $\mathcal{O}(1~\textrm{keV})$.

\begin{figure}[t]
\centering
\includegraphics[width=1\textwidth]{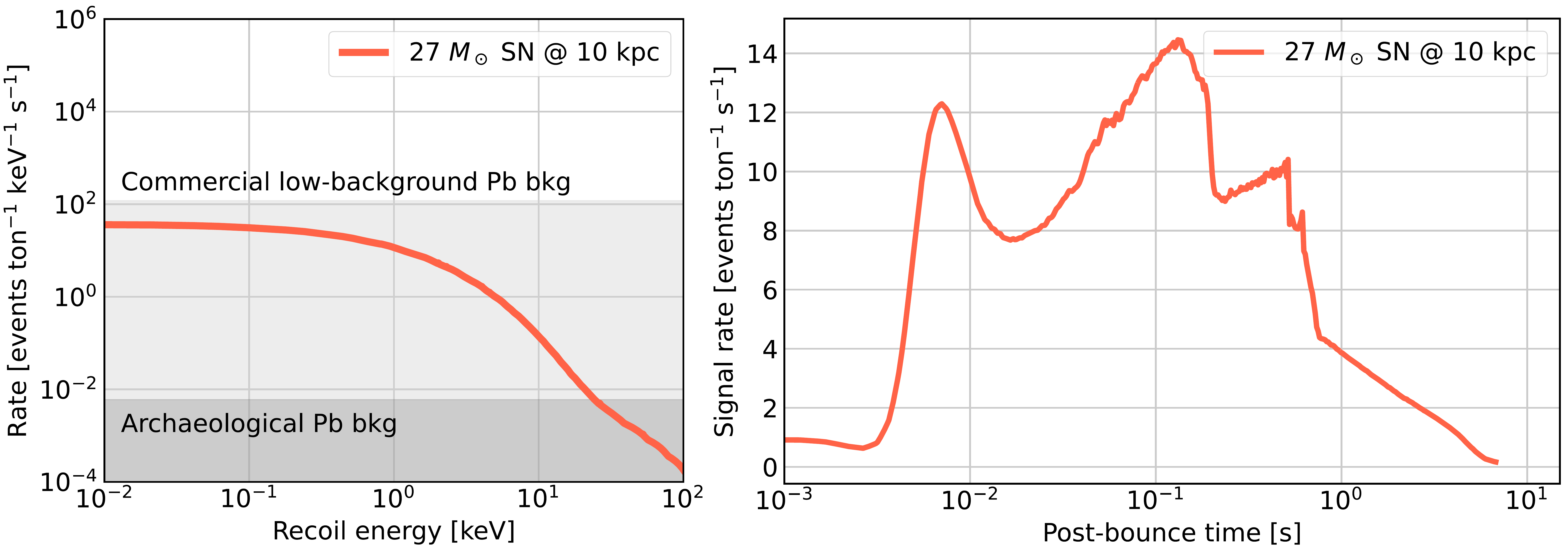}
\caption{Energy and time response of RES-NOVA to a neutrino signal produced by a core-collapse SN with progenitor mass of 27 M$_{\odot}$ occurring at 10~kpc. \textcolor{black}{The results, normalized for the detector mass, are obtained for a detector energy threshold of 1~keV and a time resolution of 100~$\mu$s, and a detection efficiency 1 above 1~keV. For details on the computation of the neutrino induced signal in the detector refer to~\cite{Pattavina:2020cqc}.}}
\label{fig:SN_RN} 
\end{figure}

The RES-NOVA experiment will implement an array of PbWO$_4$ crystals operated as cryogenic calorimeters for the detection of SN neutrinos. \textcolor{black}{The total active detector volume will be (60~cm)$^3$, equivalent to a mass of 1.7~t.} This experimental technique has the potentiality to achieve the detection of low-energy nuclear recoils with high energy resolution~\cite{Pirro:2017ecr}, as demonstrated by different experiments~\cite{CRESST_III, Agnese:2017jvy, Armengaud:2019kfj}. The expected neutrino signal in RES-NOVA, from a SN event as the one shown in Fig.~\ref{fig:SNsignal} left, is depicted in Fig.~\ref{fig:SN_RN}. There, we show the nuclear recoil energy spectrum produced by a SN event and the expected background levels when commercial low-background and archaeological Pb are employed for the crystal production. The use of archaeological Pb is therefore necessary in order to suppress the background level well below the expected signal rate. \textcolor{black}{RES-NOVA is aiming at achieving a background level of 10$^{-3}$~c/keV/ton/s in the ROI. The impact of the background can be further reduced by exploiting the geometry of the RES-NOVA detector and the topological features of SN-neutrinos events. RES-NOVA will be made of hundreds of PbWO$_4$ crystals, each of them instrumented as an independent cryogenic calorimeter. Therefore, coincidence analysis between different detectors can be implemented in order to discriminate SN-neutrinos events, which are expected to be characterized by high multiplicity (meaning that multiple SN neutrinos interact with different detectors within a given time interval), from background events, which are expected to have low multiplicity~\cite{RES-NOVA:2021gqp}.}

The characteristic time distribution of the neutrino events in the detector is also shown in Fig.~\ref{fig:SNsignal} right. The goal of RES-NOVA is to reach a detector energy threshold of 1~keV~\cite{Iachellini:2021rmh} and a time resolution of 100~$\mu$s~\cite{STRAUSS2017414}, which will allow to be sensitive to astrophysical neutrino signals. \textcolor{black}{A time resolution of 100~$\mu$s is required to identify time structures in the neutrinos emission from SN (Fig \ref{fig:SN_RN} right) and thus be sensitive to the physics of core-collapse, but also for a prompt identification of SN neutrino signals~\cite{Eller:2022ddl}.}

\subsection{Archaeological Pb-based cryogenic detectors}
\textcolor{black}{In order for RES-NOVA to be sensitive to SN neutrino signals, the implementation of archaeological Pb, with a background level compatible with the one shown in Fig.~\ref{fig:SN_RN} left, is mandatory for the detector construction.}
The background in the region of interest (RoI), namely the energy interval in which signal events are expected, must be orders of magnitude lower than the neutrino signal. Cosmic-ray interactions and natural radioactivity, namely $^{238}$U and $^{232}$Th decay chains, are responsible for the background. RES-NOVA will be installed in a deep underground laboratory, as the Gran Sasso Laboratories (LNGS, Italy), where the overburden of 3600~m.w.e. (meter water equivalent) ensures a cosmic-ray suppression of a factor ~10$^6$ compared to sea level. The major background source in PbWO$_4$ crystals is expected to come from Pb itself, in particular from $^{210}$Pb, a naturally occurring radionuclide of the $^{238}$U decay chain. Commercial low-background Pb can not be used for RES-NOVA, due to the intrinsic overwhelming concentration of $^{210}$Pb. $^{210}$Pb decays $\beta^-$ with a half-life of 22.3~y and Q-value of 63~keV, contributing with a continuum background in the RoI of SN neutrino interactions. \textcolor{black}{The RoI for these studies lies in between from the detector energy threshold to around 30~keV, depending on the type of SN emission}. Therefore, one of the most important tasks for RES-NOVA for being successful is to reduce the $^{210}$Pb content in the crystals. \textcolor{black}{Another important source of background for RES-NOVA is $^{228}$Ra, an isotope generated by the $^{232}$Th decay chain~\cite{RES-NOVA:2021gqp}. $^{228}$Ra decays $\beta^-$ with Q-value of 45.8~keV, introducing also a continuum background in the RoI.} 

Archaeological Pb offers the unique opportunity to have Pb with the lowest concentration of radionuclides. Thanks to the extremely long cool-down time, archaeological Roman and Greek Pb (about 2000 years old) is the ideal candidate. For this reason, RES-NOVA will run the first-ever detector array made of archaeological Pb-based crystals. In Tab.~\ref{tab:Pb}, we show the characteristic radioactive contaminations of commercial and archaeological (Roman origin) low-background Pb. High-purity archaeological Pb was used for the realization of the first prototypes of the RES-NOVA detector.

\begin{table}[]
\centering
\begin{tabular}{|c|c|c|}
\hline
Radionuclide & Commercial Pb & Archaeological Pb      \\
             & {[}mBq/kg{]}  & {[}mBq/kg{]}           \\ \hline
$^{232}$Th      & $<$140~\cite{Alduino:2017qet}         & $<$0.045~\cite{Alduino:2017qet}         \\ \hline
$^{238}$U       &  $<$140~\cite{Alduino:2017qet}          & $<$0.046~\cite{Alduino:2017qet}           \\ \hline
$^{210}$Pb      & 27000~\cite{OPERA:2008amt}          & $<$0.715~\cite{Pattavina:2019pxw}       \\ \hline
\end{tabular}
\caption{Main radioactive contaminations in Pb samples.}\label{tab:Pb}
\end{table}

\section{Operation and characterization of the first proof of principle RES-NOVA detector}

We operated a 840~g PbWO$_4$ crystal produced from archaeological Greek Pb~\cite{Greak_lead_Ukraine} as cryogenic detector~\cite{RES-NOVAgroupofinterest:2022pvc}. We investigated its performance and its internal radioactive contaminations. 
The crystal was housed in a highly pure Cu structure and fixed by 4 PTFE clamps, that also act as thermal link to the heat bath. The detector design followed the same one adopted in other measurements~\cite{Casali:2013zzr, Artusa:2016mat, Pattavina:2015jxe}. The detector was installed in the dilution refrigerator of the Hall-A of the underground Gran Sasso National Laboratory (Italy), and it was operated at a temperature of $\simeq$15.5~mK. The detector electronics and DAQ systems are the same one used for the CUPID-0 experiment~\cite{Azzolini:2019tta, Azzolini:2018tum}.
The crystal was equipped with a Ge-NTD thermistor~\cite{Pirro:2017ecr} as temperature sensor. This is ideal for measuring high-energy deposits in the absorber over a broad energy range, as $\alpha$-decays are expected to produce (i.e. 3-9 MeV). We focused our studies in the so called $\alpha$\textit{-region}, where we can benefit from the favourable signal-to-background ratio for the investigation of crystal radioactive contaminations. The lower $\beta$/$\gamma$-region is not considered in our radiopurity studies because of the presence of a radioactive $\gamma$ source installed next to the PbWO$_4$ crystal, that was required for the study of other detectors installed in the same experimental setup. This made impossible to retrieve any useful information about the crystal radiopurity in this energy region.

\subsection{Data analysis}
\begin{figure}[h]
\centering
     \includegraphics[width=0.48\textwidth]{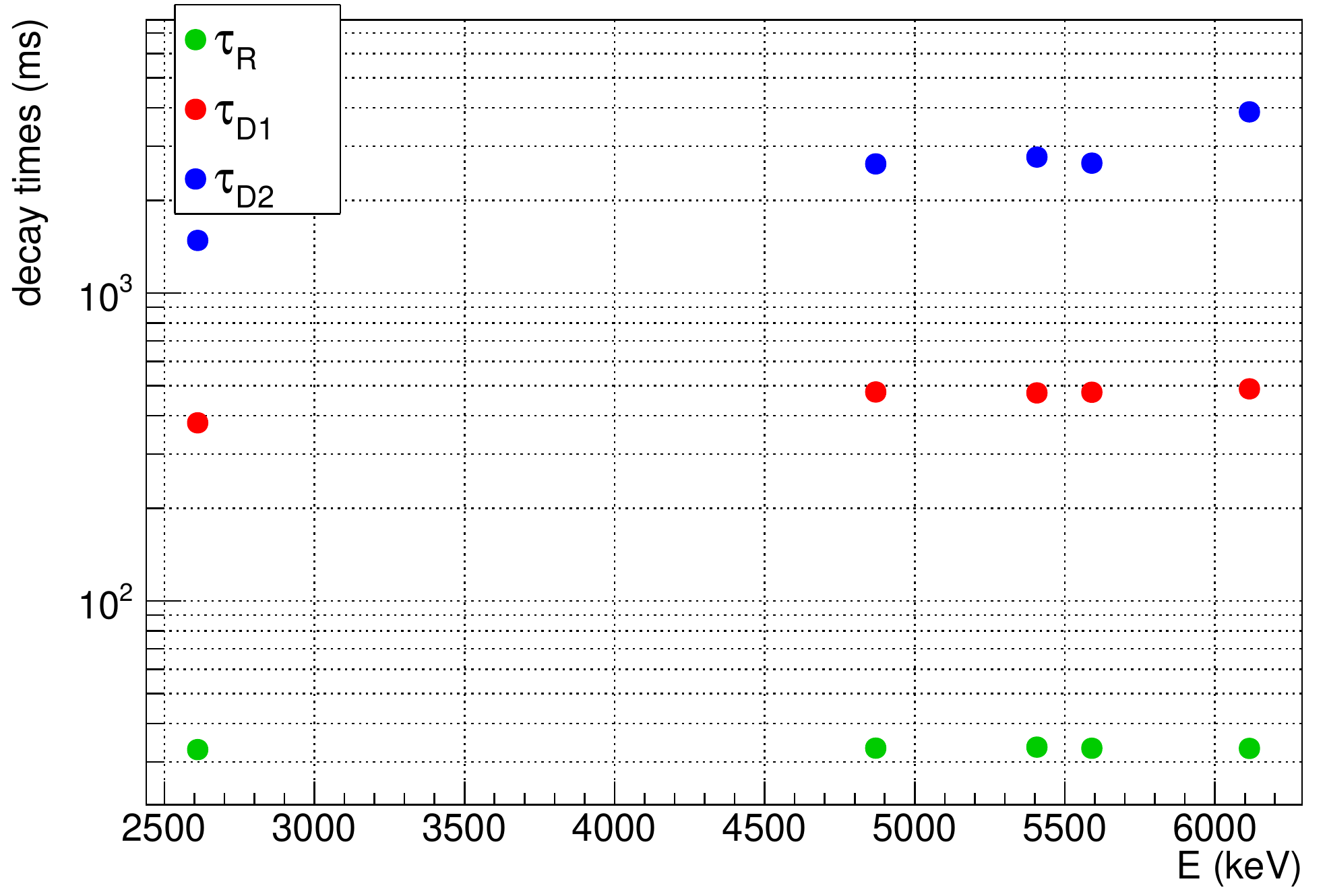}	
     \includegraphics[width=0.48\textwidth]{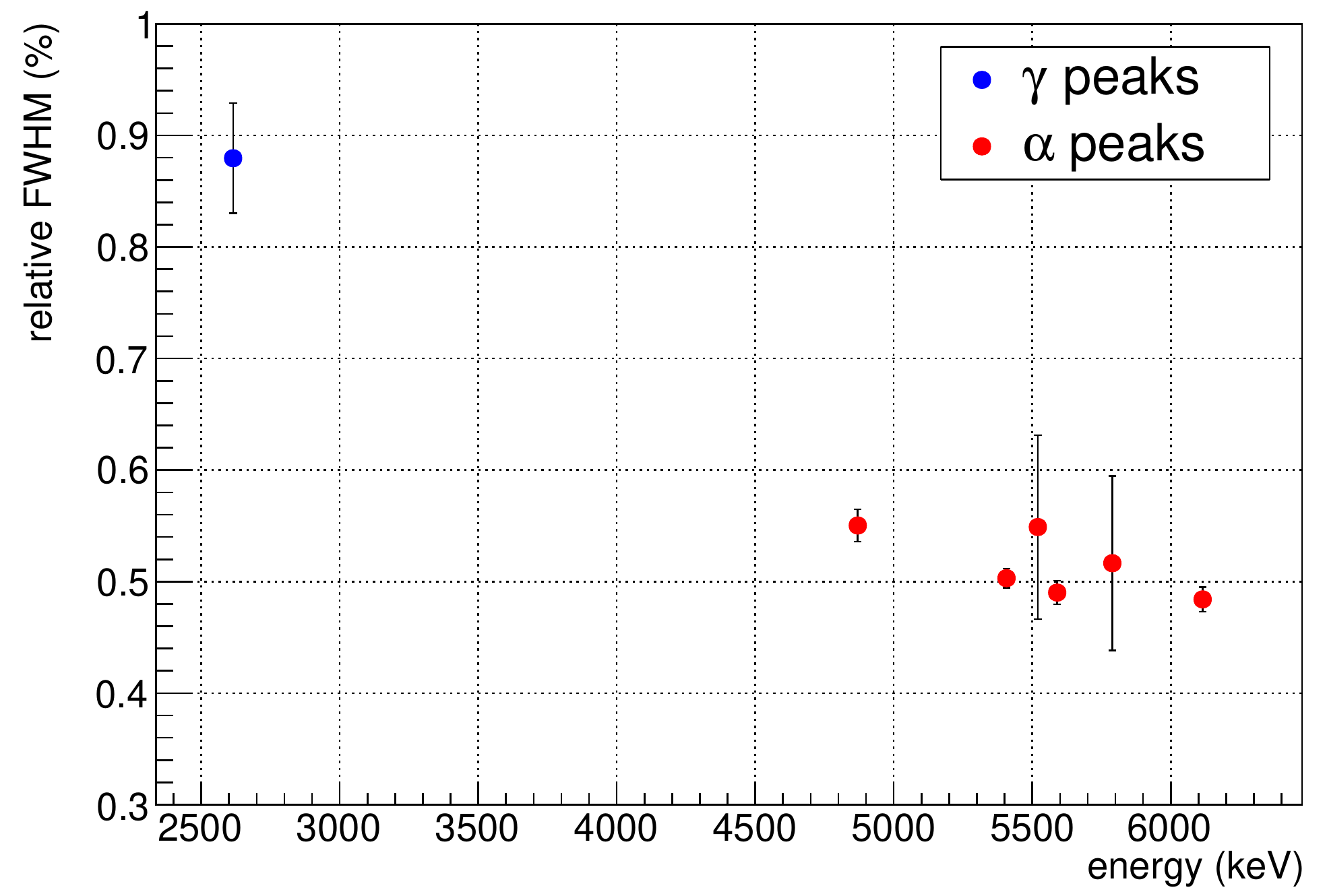}	
    \caption{\textcolor{black}{Left: Rise time and decay time constants as a function of the energy. The time constants are evaluated from the average pulses associated to signal events in $^{208}$Tl and $\alpha$ peaks. Right: Relative energy resolution of $^{208}$Tl peak and of different $\alpha$ peaks as a function of the energy.}}
    \label{fig:energy_dependance}
\end{figure}

\begin{figure}[t]
\centering
\includegraphics[width=0.5\textwidth]{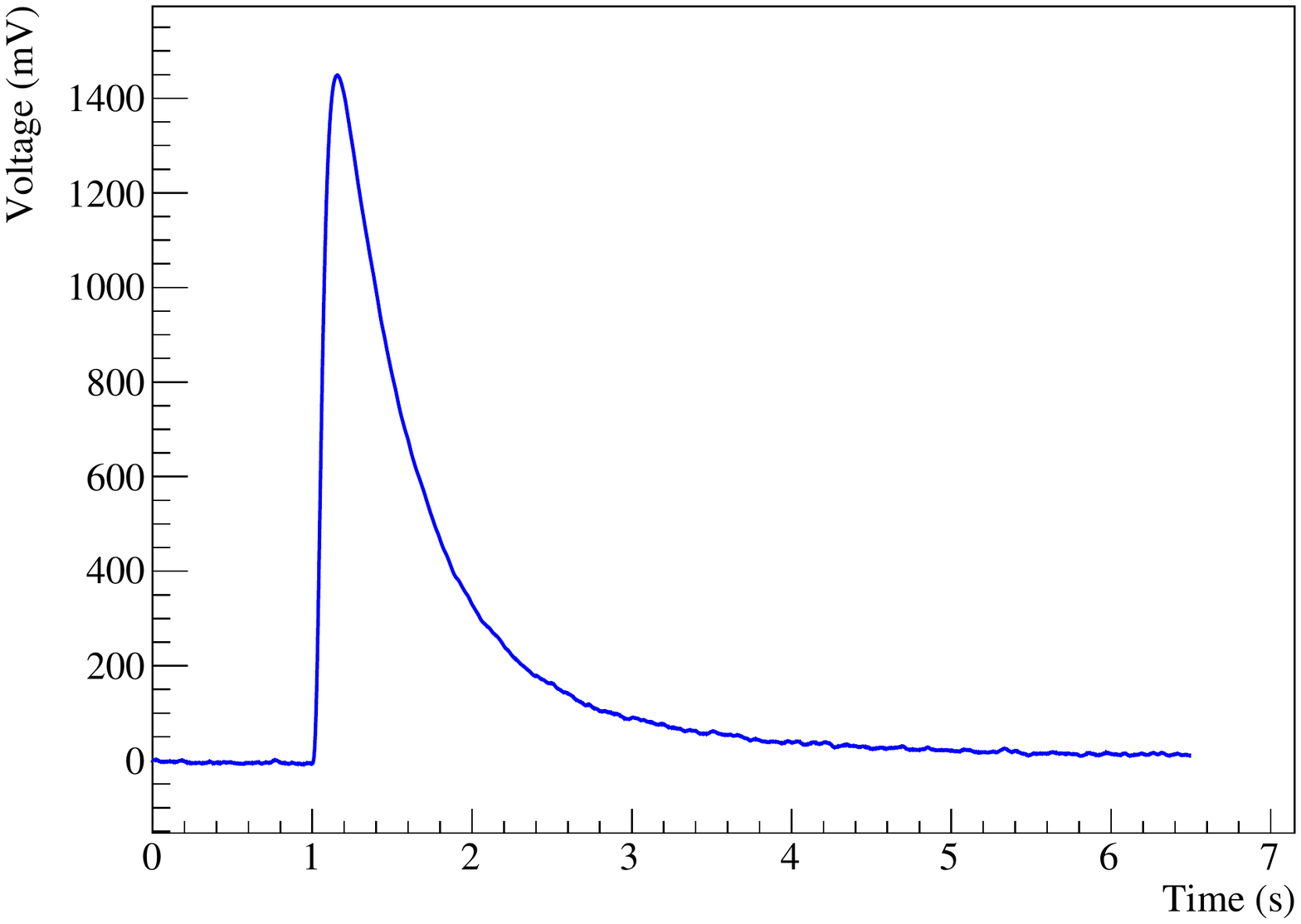}
\caption{Signal event generated by an $\alpha$ particle of 6~MeV from a $^{218}$Po radioactive decay occurring in the crystal's bulk.}
\label{fig:alpha_pulse} 
\end{figure}

The data are acquired and saved in a continuous stream, and signal pulses are identified over noise events by means of a trigger software. 
In Fig.~\ref{fig:alpha_pulse}, a typical thermal pulse generated by an energy deposition in the PbWO$_4$ crystal at $\simeq$15.5~mK is shown. The pulse develops within a time window of 7~s. An average pulse can be defined by performing a time average of single-event pulses. Assuming that the baseline noise is stochastic, the average pulse built from a large enough sample of single-event pulses is as a good approximation of the detector response (namely, of the signal pulse shape). In our detector, the pulse shape can be effectively described as the sum of three exponential functions, one of them related to the rise edge of the signal and the other two related to its decay tail. By performing the fit of the average pulse, an estimation of the characteristic rise time $\tau_R$ and of the two characteristic decay times $\tau_{D_1}$, $\tau_{D_2}$ can be obtained. \textcolor{black}{Moreover, by creating average pulses from signal events in narrow energy intervals (for example in $^{208}$Tl and $\alpha$ peaks), it is possible to study the profile of the characteristic time constants as a function of the pulses energy. Fig.~\ref{fig:energy_dependance} (Left) shows the profile of the three characteristic time constants as a function of the signal events at the PbWO$_4$ crystal operating temperature ($\simeq$15.5 mK). While the rise time is not energy-dependent, both the decay times increase at higher energies, with the slower decay time showing a stronger energy dependence.}

For the final analysis, we selected signal events whose energy was properly reconstructed. We discarded events affected by instabilities in the acquisition or affected by pile-up with other events. The cut of pile-up events is crucial: since,  due to the large mass of the PbWO$_4$ crystal, signal pulses develops over a time interval of some seconds, most of the events were affected by pile-up. Pile-up events have to be discarded since the estimation of their pulse amplitude (and consequently of their energy) would be incorrect. The only class of pile-up events that can not be discarded by analysis cuts consists in the cascade decays of $^{214}$Bi-$^{214}$Po (from the $^{238}$U decay chain). Since $^{214}$Po decays with an half-life of  $t_{1/2}$=0.16 ms, much faster that the typical time response of cryogenic detectors, the two pulses of the $^{214}$Bi-$^{214}$Po cascade are superimposed, making impossible to identify them as a pile-up event from the shape of the resulting pulse. However, such events can be clearly identified as a broad distribution in the energy spectrum at higher energies, above 8~MeV (see Fig.~\ref{fig:spectrum}).

The amplitude of the selected events was computed by implementing the Optimum Filter (OF) technique ~\cite{Gatti1986}, that is designed to maximize the signal-to-noise ratio. The resulting amplitude spectrum was converted into an energy spectrum by calibrating it on two internal contaminations $\alpha$-peaks (namely $^{210}$Po at 5.4~MeV and $^{218}$Po at 6.1~MeV) by means of a linear calibration function. Details on the analysis of cryogenic detectors data can be found in~\cite{Azzolini:2018yye}.

\textcolor{black}{After all these steps of signal selection, pile-up rejection and data processing have been applied, an overall analysis efficiency of $\simeq$ 46$\%$ was estimated. This efficiency is required in order to properly evaluate the activity of the internal contaminations of the PbWO$_4$ crystal.}

\subsection{Crystal radiopurity}
\begin{figure}[h]
\centering
\includegraphics[width=0.75\textwidth]{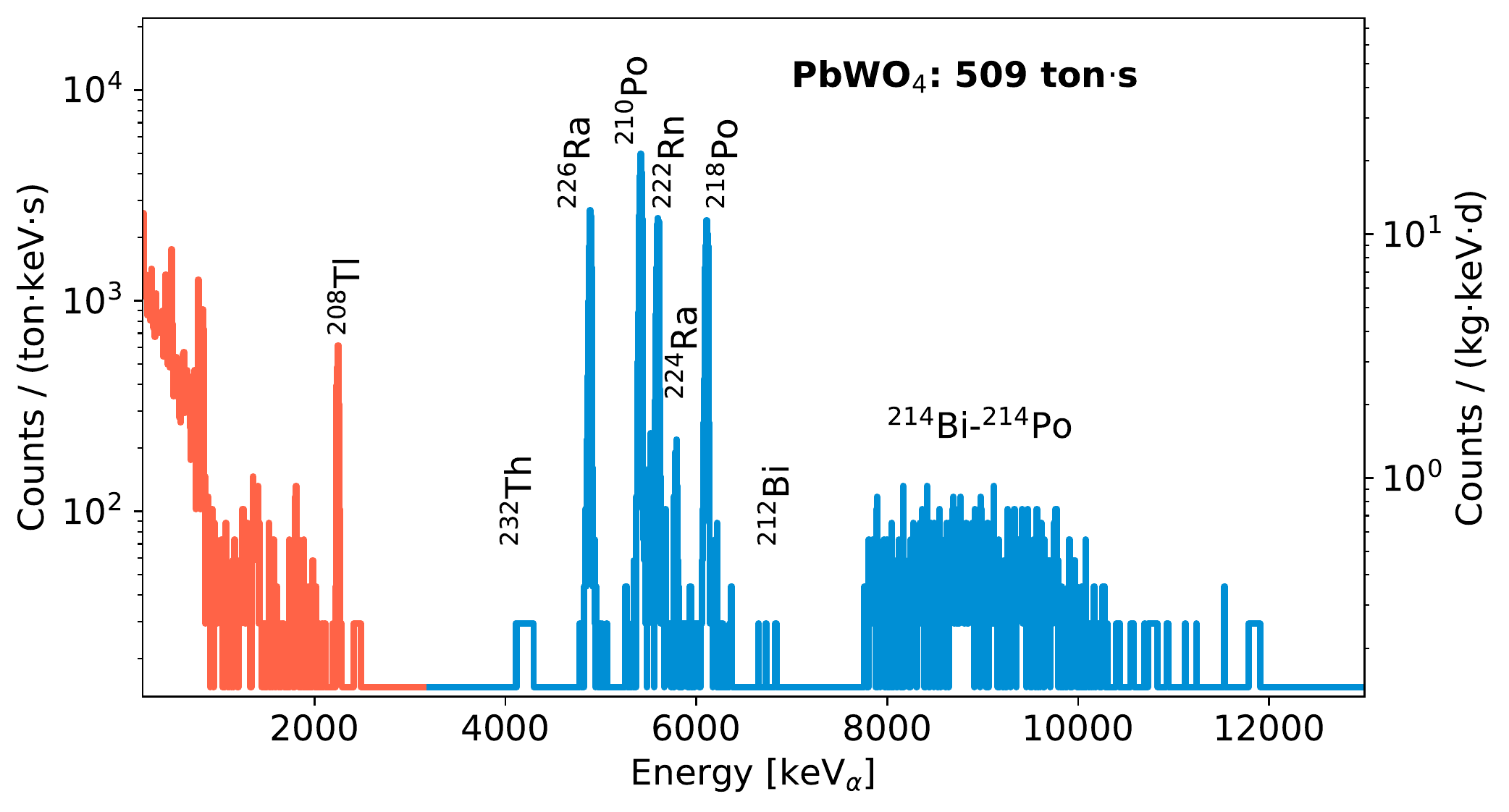}
\caption{Total energy spectrum of a 0.84~kg $^{arch}$PbWO$_4$ crystal operated as cryogenic detector. The statistics amounts to 509~ton$\cdot$s (5.9~kg$\cdot$d). The red part of the spectrum highlights $e^-$/$\gamma$ events, while the blue one represents the energy spectrum of $\alpha$ events and of $^{214}$Bi-$^{214}$Po pile-ups.}
\label{fig:spectrum} 
\end{figure}

The large mass of the crystal allowed for a high statistic study of the radioactive contaminations in the detector's bulk. Through an analysis of the $\alpha$ energy region (see Fig.~\ref{fig:spectrum}), different radionuclides were identified and their concentrations quantified. A summary of these results is shown in Tab.~\ref{tab:radio}. All the $\alpha$ decaying nuclides are ascribed to the radioactive decay chains of $^{238}$U and $^{232}$Th.

\begin{table}
\caption{Internal radioactive contaminations for a 840~g PbWO$_4$ crystal produced from archaeological Pb. Limits are at 90\% C.L.} 
\begin{center}
\begin{tabular}{lcc}
\hline\noalign{\smallskip}
Decay chain & Nuclide  & Activity \\ 
            & & [mBq/kg] \\
\noalign{\smallskip}\hline\noalign{\smallskip}
$^{232}$Th & $^{232}$Th & $<$0.04 \\
 & $^{228}$Th & 0.80$\pm$0.09 \\
 & $^{224}$Ra & 0.79$\pm$0.09 \\
\noalign{\smallskip}\hline\noalign{\smallskip}
$^{238}$U & $^{238}$U & $<$0.03 \\
& $^{234}$U & $<$0.03 \\
& $^{230}$Th & $<$0.04 \\
& $^{226}$Ra & 11.34$\pm$0.35 \\
& $^{222}$Rn & 11.60$\pm$0.35 \\
& $^{218}$Po & 11.48$\pm$0.35 \\
& $^{210}$Pb,$^{210}$Po & 22.50$\pm$0.49 \\
\noalign{\smallskip}\hline
\end{tabular}
\label{tab:radio} 
\end{center}
\end{table}

A rather small concentration of decay products of the $^{232}$Th decay chain is observed. These isotopes are not in secular equilibrium with primordial $^{232}$Th. \textcolor{black}{Possibly, the different segregation coefficients of radionuclides during the crystal growth process lead to a breaking of the secular equilibrium.} Nuclides from the lower part of this decay chain can be indirectly identified in the high energy broad distribution above 11~MeV, where pile-up of the $^{212}$Bi-$^{212}$Po $\alpha$+$\beta$ decays are expected. 
A similar behavior is also observed for the $^{238}$U decay chain, where no primordial radionuclides is detected at the level of 30~$\mu$Bq/kg, but contaminations from the lower part of the decay chain are identified. In particular, a $^{226}$Ra contamination of 11~mBq/kg  is clearly visible at 4.9~MeV. This is found to be in equilibrium with the lower part of the $^{238}$U decay chain. \textcolor{black}{In the $\alpha$ energy spectrum we can also identify a $^{210}$Po contaminations, pointing to the the presence of $^{210}$Pb. The activity of $^{210}$Pb can be assumed equal to the measured one of $^{210}$Po, since the two isotopes are in secular equilibrium. The $^{210}$Pb contaminations are generated by two different contributions:} 1) $^{226}$Ra decay from the upper part of $^{238}$U chain; 2) independent $^{210}$Pb contamination of the crystal. This second component is evaluated to be about 11~mBq/kg.

The raw materials used for the crystal production are archaeological Pb and WO$_3$. It is likely that the crystal contaminations are caused by the latter, as a result of the employment of not highly-radiopure WO$_3$. Nevertheless, we would like to point out that, this crystal is currently the PbWO$_4$ with the highest radiopurity level, ever reported in literature, with contaminations levels about three orders of magnitudes better than previous results~\cite{Beeman:2012wz}.

\subsection{Detector performance}
We present in this section the performance of the PbWO$_4$ crystal when operated as a cryogenic detector read-out by a Ge-NTD thermistor.

In Fig.~\ref{fig:spectrum} the acquired energy spectrum is shown. The energy spectrum is calibrated on two $\alpha$-peaks ($^{210}$Po, $^{218}$Po). After the spectrum calibration, the $^{208}$Tl $\gamma$-peak was reconstructed at (2245$\pm$10)~keV, a lower energy with respect to the nominal one (2615~keV). This mismatch, due to the different thermal response of the detector for $\alpha$ events and $\beta/\gamma$ events, can be used to measure the quenching factor (QF) of the detector to $\alpha$ events at the operating temperature $T$. The $QF$ is defined as:
\begin{equation}
\label{eq:QF}
QF(T) = \frac{E_{true}-E_{rec}}{E_{true}} 
\end{equation}
where $E_{true}$ is the true energy deposited in the crystal and $E_{rec}$ is the reconstructed one. In this crystal we measured  $QF(15.5~mK)$=(14.2$\pm$0.9)$\%$, which is slightly lower than the value observed in~\cite{Beeman:2012wz}.

An important parameter for cryogenic detectors is the baseline resolution, since it quantifies the impact of thermal, electronic and vibrational noise on the baseline fluctuations. The baseline resolution determines the low-energy threshold of the detector and it contributes to its total energy resolution. The baseline resolution of this detector, evaluated as the FWHM of the energy distribution of baseline traces where no trigger was fired, is FWHM$_{noise}$=(8.8$\pm$0.2)~keV. This energy resolution does not match the requirements of RES-NOVA. In fact for the final experiment the plan is to use Transition Edge Sensors (TES)~\cite{Pirro:2017ecr} instead of Ge-NTDs, since TES grants better performances.

We have also evaluated the total detector energy resolution over the entire $\alpha$-region (in the range 4-7 MeV). The average energy resolution of $\alpha$ peaks is FWHM$_{\alpha}$=(28.5$\pm$1.4)~keV, corresponding to a relative energy resolution of (0.52$\pm$0.03)$\%$. The energy resolution of $^{208}$Tl $\gamma$-peak was estimated to be FWHM$_{\gamma}$=(23.0$\pm$1.5)~keV, corresponding to a relative energy resolution of (0.87$\pm$0.06)$\%$. \textcolor{black}{Fig.~\ref{fig:energy_dependance} (Right) shows the profile of the relative energy resolution as a function of the energy of $^{208}$Tl peak and of different $\alpha$ peaks.} All these values, which concur with the ones achieved by other massive cryogenic detectors~\cite{CCVR}, show the potential of the cryogenic experimental technique. 

We performed also a charaterization of the detector sensitivity, $S_c$. This is a temperature-dependent parameter that quantifies the conversion between the energy released in the crystal $E$ and the amplitude of the electric pulse $A$ normalized for the electronic gain $G$ of the acquisition system. It is defined as:
\begin{equation}
\label{eq:crystal_sensitivity}
S_c(T) = \frac{A}{G E}
\end{equation}
The sensitivity for this detector was evaluated by selecting signal events within the $\alpha$-region of the energy spectrum. The measured value at 15.5 mK is $S_c$=(46.25$\pm$0.24)~$\mu$V/MeV. This value is comparable with other detector benchmark as the TeO$_2$ crystal operated by the CUORE cryogenic experiment~\cite{CUORE:2016aqq}.

\section{Conclusions and outlook}
In this work, we presented the characterization of a large mass PbWO$_4$ crystal, produced from archaeological Pb, operated as cryogenic particle detector. The detector showed an excellent energy resolution below 1\% over a broad energy range between 2.6-7~MeV. This value concurs with other well established cryogenic detectors (e.g. TeO$_2$). The crystal, thanks to the employment of archaeological Pb, features a high radiopurity level, about 3 orders of magnitude better than reported in literature for the same compound. However, some concentration of radioactive impurities is still detected. The activities of $^{226}$Ra, $^{228}$Th and $^{210}$Pb are measured at the level of a few mBq/kg, and these are ascribed to possible impurities in the WO$_3$ used for the synthesis of the PbWO$_4$ powder. We are planing to operate other PbWO$_4$ crystals produced from raw materials with higher purity level, aiming at assessing this excess of contamination.
This work represents a major milestone for the RES-NOVA project, which is aiming at operating an array of ultra-low background PbWO$_4$ cryogenic detectors for the investigation of astrophysical neutrino sources. The results showed in this manuscript demonstate the feasibility of this detection technology. However, a further improvement of the crystal radiopurity is needed, with dedicated screening campaigns and additional purification of the raw materials employed for the crystal production.

\section{Acknowledgments}
This research was partially supported by the Excellence Cluster ORIGINS which is funded by the Deutsche Forschungsgemeinschaft (DFG, German Research Foundation) under Germany’s Excellence Strategy - EXC-2094 - 390783311. F.A. Danevich and V.I. Tretyak were supported in part by the National Research Foundation of Ukraine Grant no. 2020.02/0011.


 \bibliographystyle{elsarticle-num} 
 \bibliography{elsarticle-template-num}





\end{document}